Title:

Ion-beam sputtering of NiO hole transporting layers for p-i-n halide perovskite solar cells

Authors:

P. Gostishchev[1#], L.O. Luchnikov[1#], O. Bronnikov[1], V. Kurichenko[1], D.S. Muratov[1], A.A. Aleksandrov[2], A.R. Tameev[2], M.P. Tyukhova, S. Le Badurin. I.V.[1,3], Ryabtseva M.V.[3], D. Saranin[1*] and A. Di Carlo[4*].

Affiliation:

[1]LASE – Laboratory of Advanced Solar Energy, NUST MISiS, 119049 Moscow, Russia

[2] Laboratory "Electronic and photon processes in polymer nanomaterials", Russian Academy of Sciences A.N. Frumkin Institute of Physical chemistry and Electrochemistry, 119071, Moscow, Russia

[3]Department of semiconductor electronics and device physics, NUST MISiS, 119049 Moscow, Russia

[4]CHOSE – Centre of Hybrid and Organic Solar Energy, Department of Electronics Engineering, Rome, Italy

[#] *The authors contributed equally*

*Corresponding authors:*

*Dr. D. Saranin email: saranin.ds@misis.ru*

*Prof. Aldo Di Carlo: aldo.dicarlo@uniroma2.it*

Abstract:

Ion-beam sputtering offers significant benefits in terms of deposition uniformity and pinhole-free thin-films without limiting the scalability of the process. In this work, the reactive ion-beam sputtering of nickel oxide has been developed for the hole transporting layer of a p-i-n perovskite solar cells (PCSs). The process is carried out by oxidation of the scattered Ni particles with additional post-treatment annealing regimes. Using deposition rate of 1.2 nm/min allowed growth of very uniform NiO coating with the roughness below 0.5 nm on polished Si wafer (15x15 cm$^2$). We performed a complex investigation of structural, optical, surface and electrical properties of the NiO thin-films. The post-treatment annealing (150-300 ˚C) was considered as an essential process for improvement of the optical transparency, decrease of defects concentration and gain of the charge carrier mobility. As result, the annealed ion-beam sputtered NiO films delivered a power conversion efficiency (PCE) up to 20.14%, while device without post-treatment reached the value of 11.84%. The improvement of the output performance originated from an increase of the short-circuit current density ($J_{sc}$), open circuit voltage ($V_{oc}$), shunt and contact properties in the devices. We also demonstrate that the ion-beam sputtering of NiO can be successfully implemented for the fabrication of large area modules (54.5 cm$^2$) and PSCs on a flexible plastic substrate (125 microns).

Introduction:

Halide perovskites (HPs) are a promising material for the next generation of photovoltaics (PV) due to their remarkable optical and transport properties[1]. This makes HPs suitable for use in high-performance photovoltaic devices with power conversion efficiencies (PCEs) up to 25.8%[2]. The fabrication process of perovskite solar cells (PCSs) has been developed in the last year including several large-scale solution coating methods (slot-die[3], doctor blade[4], etc.), which are cost-effective for the mass production with low capital expenditures (CAPEX)[5]. However, it is hard to avoid completely the use of physical (usually vacuum based) processes in the full cycle of HP PV fabrication. In general, HP-based solar cells are thin-film devices that consist of a sub-micron thick perovskite absorber sandwiched between p- and n-type charge transporting layers. The electrodes of thin-film devices typically are transparent conductive oxides[6] (TCOs) - like $In_2O_3:SnO_2$, $ZnO:Al$ - and metal films[7] (copper, aluminum, silver and gold). The TCO coatings are used as a front electrode on the transparent substrates (glass or plastic), which allow the penetration of the light to the absorber film. The metal films are used as a back electrode and fabricated with physical vapor deposition based on thermal evaporation process. The fabrication of TCO, based on the wide band-gap semiconductors, is already widespread in the industry of the organic light-emitting diodes, thin-film optical sensors, etc[8]. The most common approach for mass production consistes on the use of magnetron sputtering (MS)[9]. In this method, a gaseous ions accelerated from the plasma under magnetic field, bombard the target and induce the non-thermal physical ejection from the surface. Using the billets of ITO on glass or plastics is a standard process for prototyping of PSCs and upscaled modules. Moreover, the sputtering technique could be effectively applied to charge transporting layers (CTLs) based on the oxide materials, for example nickel oxide (NiO)[10], molybdenium oxide ($MO_3$)[11] for p-type transport, and tin oxide ($SnO_2$)[12], zinc oxide (ZnO)[13] for n-type transport. Therefore, the deposition of the multilayer stacks with TCO electrodes and oxide CTLs could be realized in the closed loop of the manufacturing. The fabrication of charge transporting layers could be done with sputtering from ceramic oxide targets, or via reactive sputtering with presence of the gaseous oxygen during the deposition. However, the MS processing has several critical features – the direct interaction of the plasma and substrate could be accompanied with formation of the discharge arcs, which induce the damaging of the growing film[14]. Recently, *Reddy et. al.*[15] presented a complex investigation for the critical parameters in damage-less MS of barrier films in p-i-n PSCs.

Alternatively, the ion-beam sputtering (IBS) can be applied to the deposition of the oxide CTLs. IBS technique utilizes a high-energy ion beam to remove the material from a target and deposit it onto a substrate. In contrast to the MS processing, the IBS deposition provides the separation of ion generation, sputtering, and coating growth[16] with the ion-beam source facing the target. After interaction of the ion-beam with the surface of the target, the sputtered particles will deposit on the substrate, permitting the film to grow. The ion energy can also be adjusted (800-5000 eV)[17] to optimize the sputtering rate, enabling the deposition of thick films in a relatively short time. IBS has a low thermal impact on the substrate and minimal contamination, as the ion beam is less likely to generate particles compared to other sputtering methods. This results in improved film quality and a lower risk of substrate damage, especially for temperature-sensitive substrates. IBS can produce films with high uniformity and reduced surface roughness[18]. This is because the ion beam is able to more effectively sputter the target material, resulting in fewer defects and impurities in the deposited film. So far such technique has not considered for HP PVs[19], [20]. Several reports show the benefits of the ion-beam sputtering for damage-free deposition of TCO in organic-light emitting diodes[21], electrodes on plastics[22], and anti-reflection coatings for III-V optoelectronics[23].

In this paper, we present a detailed investigation on the application of the ion-beam sputtering method for the fabrication of the charge-transporting layers in PCSs. Our results show that reactive ion-beam sputtering of nickel oxide from the metal target can be used as for high-performing hole transporting layers in devices with p-i-n structure. We found that the annealing post- treatment has a critical impact on the structural, surface and optical properties of the NiO thin-films after deposition. Our findings demonstrate the high potential of the ion beam sputtering for the fabrication and up-scaling of the perovskite solar cells.

**Results and discussions**

The nickel oxide hole transporting layer (HTL) is deposited via ion-beam sputtering with a base pressure in the chamber of $4*10^{-2}$ Pa. The general scheme of the ion-beam deposition chamber together with the structure of PSC is depicted in the **fig.1**. The 200 mm wide ion-beam source (prolonged ion source with closed electron drift [24]) was slightly tilted toward the horizontal plane. The target holder was mounted with a tilt of 40 degrees to provide optimal direction of the sputtered material to the substrate holder, as it shown on the **fig.1(a)**. The substrates for the deposition of NiO thin films were mounted onto an auto-sliding metal plate. We used Ni metal target (200 x 150 mm$^2$) with a purity of 99.999% for the reactive conversion to NiO. The oxidation of the sputtered Ni was provided by O$_2$ flow (2.2 cm$^3$/min) directed to the substrate holder. The argon (99.998% purity) was used as a working gas for the generation of the ion-beam. We used constant parameters of the ion-beam source with the voltage of 1000 V and current of 100 mA, which enabled the deposition rate of 1.2 nm per minute. Prior to the process, the Ni target was etched with an ion beam for 10 minutes to remove the oxidized surface layer and possible impurities. In this work, we fabricated p-i-n perovskite solar cells accord with the multilayer stack presented on the **fig.2(b)**. The NiO HTL was sputtered on the top of pre-patterned ITO-glass substrates, the details for the manufacturing of perovskite absorber Cs$_{0.2}$CH$_3$(NH$_2$)$_2$PbI$_3$ (CsFAPbI$_3$), C$_{60}$/Bathocuproine (BCP)–electron transport layers and copper cathode described in the experimental part.

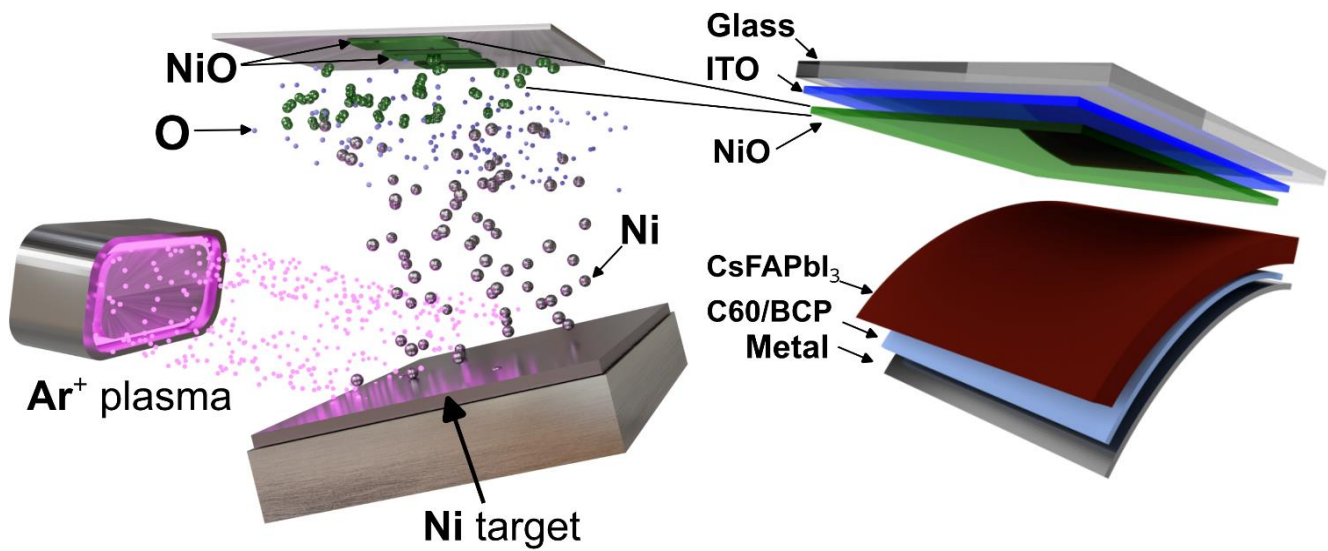

Figure 1 – The general scheme of ion-beam sputtering process (left) and device structure (right) of the perovskite solar cell with NiO HTLs

The thickness uniformity of the 7 nm NiO was assessed by ellipsometric (Senresearch 4.0 from SENTECH) measurements across the polished Si substrate (Diameter – 15 cm), as shown on the **fig.2**. The analysis of ellipsometry data demonstrated the average thickness value =6.9 nm with negligible standard deviation (±0.11 nm or 1.5%) over the substrate. This illustrates the high potential of the ion-beam sputtering process for the fabrication of planar and uniform coatings of a large area.

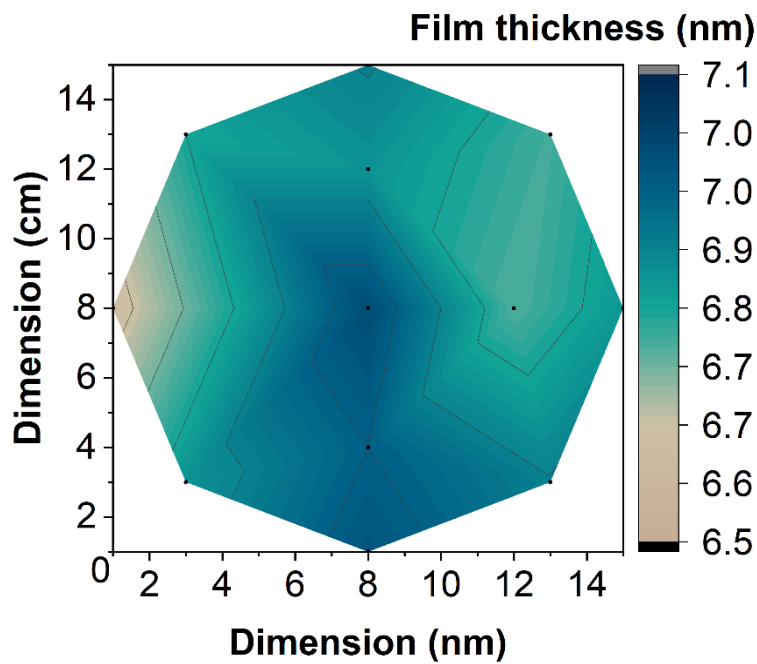

Figure 2 – The thickness uniformity of the ion-beam sputtered NiO film on polished Si substrate

Structural, optical, and electrical properties of NiO films strongly depend on the post-deposition treatment conditions[25]. The air annealing procedure was considered as one of the effective approaches to control the microstructure, stoichiometry, and transport properties of nickel oxide layers for optoelectronic devices [26]. In order to investigate the impact of the annealing on the deposited NiO properties, we performed the comparative analysis of NiO thin-films post-treated at different temperature regimes. The top limit was fixed at 300 ˚C corresponding to the loss of conductivity in ITO-glass substrates[27] used for the manufacturing of PSCs. The ion-beam sputtered NiO thin-films with post-treatment annealing in the ambient conditions are listed in the following: not annealed (**NA**); annealed for 60 minutes at 150 ˚C (**150C**); annealed for 60 minutes at 300 ˚C (**300C**). To simplify the naming of the thin-films and devices in the manuscript in the following the configuration of the NiO HTL will be identified with the corresponding annealing temperature.

**Figure** 3 (c, d, e) shows Ni2p spectra of IBS NiO **NA**, **150C** and **300C** samples. The Ni2p line of NiO$_x$ can be identified by the unique NiO multiplet splining [28, 29] with additional intensity increment at 854.9 eV, which indicates presence of NiOOH, NiOH$_2$, defective NiO or 'Ni$_2$O$_3$'. As it was previously demonstrated in studies with similar preparation methods of NiO$_x$ films, oxygen in vacuum as reactive gas results in Ni-O film without Ni(OH)$_2$ or NiOOH contamination [31, 33]. By adding pairs of multiplets NiO + NiOOH or NiO + Ni(OH)$_2$ in Ni2p spectra we did not achieve a good match with the experimental curve (Fig. S2). For defective NiO and 'Ni$_2$O$_3$' no good multiplet function is suggested yet. In our model the residual area that cannot be fitted by NiO multiplet was enveloped by four Gauss-Lorentz functions and was attributed to so called 'Ni$_2$O$_3$' state, represented by defective nickel oxide with an excess of oxygen. In Ni2p spectra fraction of NiO in IBS NiO$_x$ films grows from 77.7 and 76.6 % for **NA** and **150C** sample to 88.4 % for **300C**. The O1s

spectrum of NiO$_x$ (**fig. S1**) comprises of three peaks, a carbon oxygen species (C-O, C-O-O, C-O-H) from organic surface contaminant at 532.6 eV, low binding energy feature (LBE) at 529.1 eV which is assigned to the lattice oxygen in NiO$_x$ and high binding energy feature (HBE) at 530.9 eV. The high binding energy feature is discussed in literature and arises due to oxygen defects, adsorbed oxygen, or spontaneous hydroxylation of the Ni surface [32, 33]. We prefer to assign this feature to 'Ni$_2$O$_3$.'. In our case, the hypothesis of defective nickel oxide existing in the obtained films is based on the following facts: firstly, the XRD peaks of the nickel oxide film are shifted towards smaller diffraction angles, which indicates the lattice spacing widening, due to the presence of embedding oxygen atoms caused by the IBS process. The XRD peaks of nickel oxide (**fig. 3a**) shift to the right from 64.66° to 65.51° 2θ degrees as 300 °C annealing applied, which indicates lattice relaxation and its approximation to stoichiometric composition. Secondly, the effect of a significant increase in the nickel oxide film transparency (**fig. 4 b**) after 300 °C annealing is observed. Thirdly, from the XPS survey spectra (**fig S3, table S1**), it is observed that the relative concentration of oxygen in the NiO$_x$ film decreases from 1.15 to 1.08 at higher annealing temperature. These observations allow to conclude that during the IBS production of nickel oxide an excess (above stoichiometric) of oxygen entered the interstitials of the lattice, thereby expanding the volume of NiO unit cell (up to 78.9425 Å$^3$ for **NA** sample) [31] and creating nickel atoms with an altered oxidation state and oxygen interstitials.

IBS Nickel oxide films morphology obtained via atomic force (**fig.** S4) demonstrate almost twofold higher surface roughness coefficient (R$_{sq}$) from 0.84 to 1.48 nm for NA and **150C** samples. The value of R$_{sq}$ for the films annealed at 300 °C is slightly lower if compared to the films annealed at 150 °C. Mean grain size calculated from AFM data (fig. 3 f, g, h) is also higher for **150C** sample, compared to the not annealed ones (13.6 and 9.8 nm respectively), while 300C sample has slightly bigger grains as well. An increase in grain size can be explained by surface recristallization at higher annealing temperature reaching an equilibrium state which is also confirmed by FWHM of the first two peaks on the XRD pattern (fig. 3 a).

By elevating the temperature during annealing, the excess oxygen within the NiOx film is thermally desorbed, leading to an increase in the stoichiometric ratio of NiO. This process also results in a modification of the NiOx surface energy. The water contact angle of the NiOx films (**Fig.** 3b) was characterized using a KRUSS drop shape analyzer DSA30S, with results indicating that the contact angle remained unchanged with an increase in grain size and roughness for the 150C sample (**Fig.** 3 f, g, h). However, for the 300C sample, a significant decrease in the contact angle was observed, with a value of 26° (**Fig.** 3b). This reduction in contact angle indicates a greater contribution of polar interactions to the surface energy of annealed films.

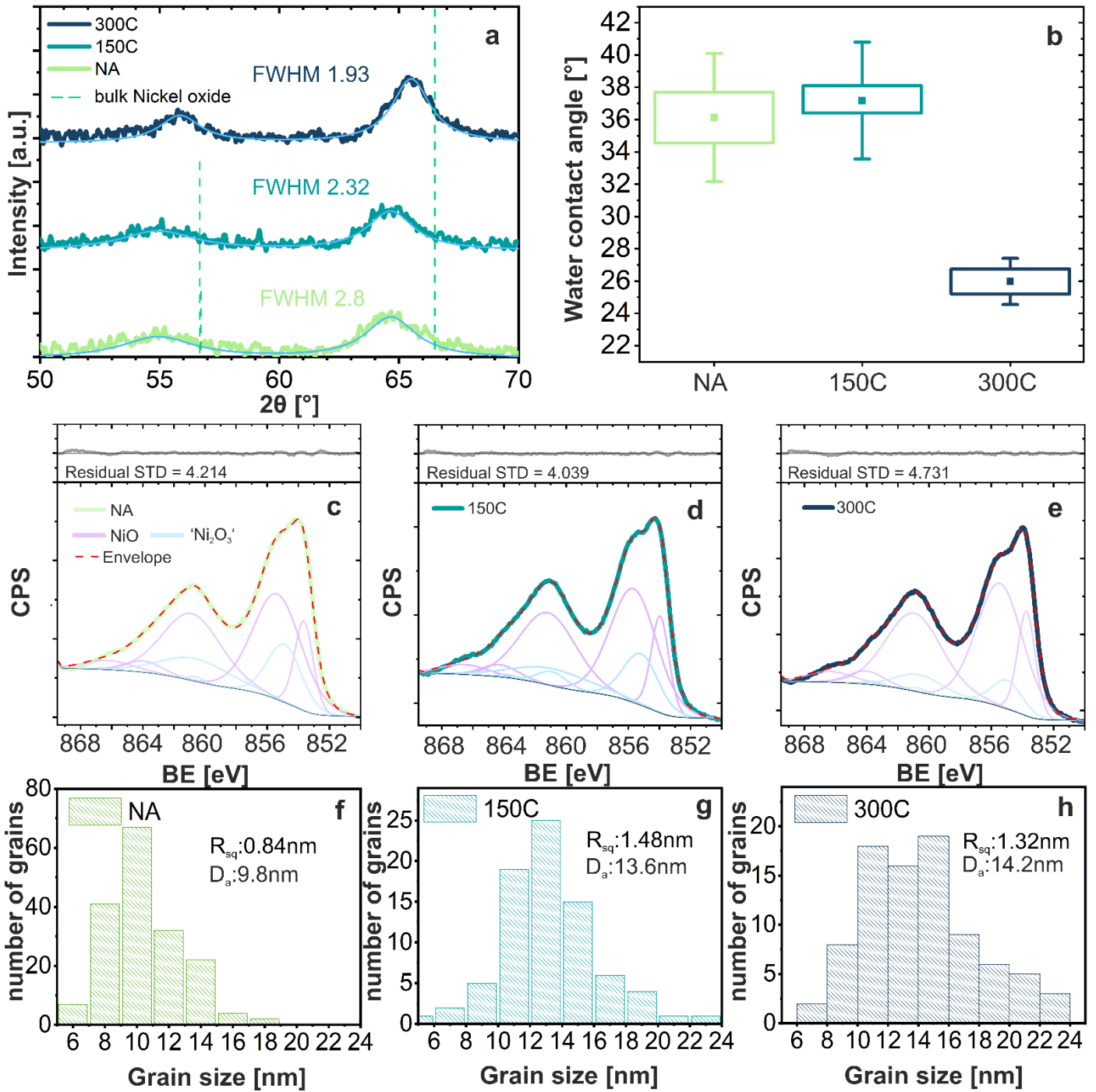

Figure 3 – The X-ray diffraction patterns of ion-beam sputtered NiO thin-films (a); the measured values of the wetting angles for **NA**, **150C** and **300C** NiO samples (b); the X-ray photoelectron spectra of the **NA** (c), **150C** (d) and **300C** (e) NiO thin-films; the results of AFM analysis with histograms of the grain size distribution (f)-(h)

Considering the direct band gap properties of nickel oxide[28], [29], we used the Tauc plot equation[30] **(eq.S1 in S.I.)** to estimate the band-gap values ($E_g$) of the fabricated thin-films **(fig. 4 (a))**. The $E_g$ values extracted through the linear extrapolation of the plot is 3.280 eV for NA, 3.459 eV for 150C and 3.506 eV for 300C, which is in good agreement with data reported in the literature[31]–[34]. The transmittance data presented in **Figure 4(b)** reveals a pronounced impact of parasitic absorbance within the 350-450 nm range for NA NiO, with transmittance values ranging from 75-85%, compared to the nearly 95% transmittance observed for the quartz substrate. The transmittance of NA NiO films exhibits a slight increase for longer wavelengths (reaching 87% at 550 nm and 90% at 900 nm). In contrast, both 150C and 300C NiO films demonstrate improved transparency, with transmittance values ranging from 75-88% within the 350-450 nm range and 92.5% at 900 nm. Notably, the transmittance spectra of the 300C NiO layer is similar to that of the quartz substrate for wavelengths larger than 650 nm, indicating the absence of optical absorption in this region. The 150C NiO exhibits a slightly lower transmittance than the 300C NiO, with a reduction of approximately 1-2% within the 350-780 nm wavelengths.

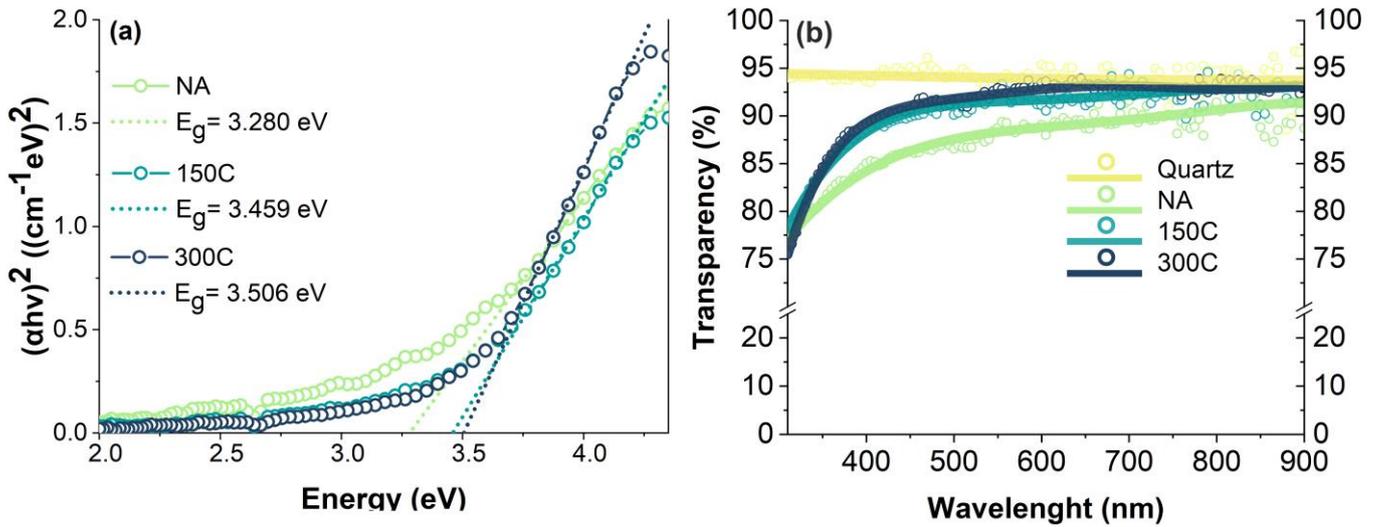

Figure 4 – The Tauc plot with calculation of the $E_g$ values from the absorption spectra (a), the transmission spectra of the NiO films deposited on the quartz substrates (b)

The electrical properties of the NiO thin-films were estimated from measurements of specific resistivity and hole mobility values (**tab.1**). The electrical conductivity of NiOx films was measured using the four-probe method, while the mobility of charge carriers in NiOx films was measured based on the Hall effect in a weak magnetic field (see **S.I.** for the **fig.SS5, S6**, **tab.S3,S4** and **eq.S1-S7**).

The specific resistivity ($\rho$) of the films decreased from ~$10^{-4}$ Ohm*cm for NA and 150C NiO to ~$10^{-5}$ Ohm*cm for 300C. The reduction of $\rho$ is due to improved charge carrier mobility. The Hall mobility measurements demonstrates that NiO films have p-type conduction. The Hall mobility for NA sample is 4.05 $cm^2V^{-1}s^{-1}$ and slightly decreases (1.98 $cm^2V^{-1}s^{-1}$) by annealing at 150˚C. However, increasing the annealing temperature to 300 °C we observe a significantly increases the Hall mobility up to 52.56 $cm^2V^{-1}s^{-1}$. The electrical properties of nanocrystalline NiO films have a strong dependence on point defects of the crystal, such as nickel vacancies($V_{Ni}$) [35] and interstitial oxygen($O_i$)[36]. In our work, we used the reactive sputtering methods for the fabrication with different annealing regimes in the oxidation atmosphere (air), thus the post-treatment can induces changes of the defect types. As reported[36], [37], the additional introduction of the oxygen to NiO films (in this work–air annealing) provides partial ionization of the Ni with a raise of hole concentration in the material and associated with improved conductivity. *Poulain et al.*[38] demonstrated that

NiO deposited via room-temperature reactive sputtering has oxygen rich grain boundaries, that provides electrical conduction through trapped charges. Moreover, defects (oxygen interstitials) and the associated charge transport along the grain boundaries in nanocrystalline NiO thin-films can be suppressed with an annealing step above 200 °C. According to our results, we could assume that the improved charge carrier mobility and conductivity for 300C samples is caused by an improved quality of the nanocrystallites (enlarged grain size) and a decrease in the $O_i$ concentration. This also supports with an observation that 150C post-treatment is not enough to improve the charge transport in the NiO thin films.

Table 1. The electric and transport parameters extracted for NiO thin-films

| Device configuration | Specific resistivity $\rho$, Ohm·cm | Hole mobility $\mu_H$, cm$^2$V$^{-1}$s$^{-1}$ |
|---|---|---|
| NA | 0.23*10$^{-4}$ | 4.05 |
| 150C | 0.17*10$^{-4}$ | 1.98 |
| 300C | 0.9*10$^{-5}$ | 52.56 |

Three distinct device sets were fabricated, taking into account the investigated annealing temperatures of NiO thin films. The photoelectrical parameters extracted by current-voltage (IV) characteristics (AM1.5 G conditions), are reported on the **fig. 5(a)** for champion devices while the statistical distribution of measured values is reported in **fig.S1** (**Supplementary information – S.I.**). The NA PSCs showed poor performance with the best PCE of 11.84%, while post treated devices demonstrated an improved PCE up to 15.58% for 150C PSC and up to 20.14% for 300C. The annealing of ion-beam sputtered NiO at 150˚C and 300˚C allowed to achieve performances similar to those obtained with vacuum processed HTLs [39]–[42]. Notably, the observed PCE improvement mostly originated from the increase in short-circuit current density ($J_{sc}$). Concerning the NA PCSs, the average increase of $J_{sc}$ for 150C and 300C devices was +22% and +35%, respectively. The values of open-circuit voltage measured for the fabricated PSCs are also correlated with the post-treatment conditions. The statistical distribution of the output IV parameters presented in the box-charts on the **fig.S5** in **S.I**. The highest average $V_{oc}$=1.01 V was reached by 300C PSCs. interestingly, the 150C device showed the lowest average $V_{oc}$=0.91 V, while NA PSCs demonstrated sufficiently higher values of 0.98 V. In general, the $V_{oc}$ is related to the splitting of quasi-Fermi levels (QFLs)[43] in the absorber under light exposure and the energy band-gap of the absorber. Typically, the real values of $V_{oc}$ measured for solar cells are smaller than theoretically predicted due to the thermalization losses[44] and imperfections of the device structure. The mismatch between theoretical values of $V_{oc}$ and experimental data is explained by energy level offsets at the semiconductor junction of HP absorber with charge-transporting layers and results in bending of QFLs. At the same time, the interface quality of HTLs (surface traps) could give significant impact on the non-radiative recombination processes and negatively affect to the charge collection efficiency[45].

Thus, the difference of $V_{oc}$ for the investigated PSCs could be attributed to a combination of the optoelectronic and surface properties of NiO. We should assert the point that competitive performance of PSCs with IBS processed HTL was achieved for an intrinsic type of the material without any additional passivation[46], doping[47] or interface engineering[48], typically used for NiO. The values of $V_{oc}$ obtained for our structures (<1.1 V) show that devices have the voltage losses. Potentially, this feature could be improved via doping of NiO[49], [50] (using Ni-Cu or Ni-Co alloy targets) or interface passivation[51] (deposition of the wide band-gap ultra-thin interlayer, such as MgO). Both approaches could be easily implemented with an ion-beam sputtering process without limitation for scalability.

The filling factor (FF) values of PSCs with different post-treatment of ion-beam sputtered HTL showed a straight correlation to the annealing temperature. The average value of FF for NA devices was 71%, which slightly increased to 73% with annealing at 300 ˚C.

As confirmed by external quantum efficiency spectra on the **fig.5(b),** the $J_{sc}$ values extracted after integration are showing remarkable improvement for devices with annealed NiO HTL. The comparison of EQE spectra for NA and 150C PSCs shows the clear impact of the improved transmittance for the annealed device in the 360-700 nm region and verifies the optical data from the **fig.4(b)**. A close inspection of the plot shows that the increase in the spectral performance takes place only for short and middle wavelengths, while for 700-800 nm region the EQE values are almost the same. For 300C PSC we observe an additional gain of photon to current conversion with respect to 150C, in all the spectral range of the absorbance with peak values of 92% at 410 nm and 87% at 600 nm. Maximum power density measurements (**fig.5(c)**) for 150C and 300C PSCs show a stable value without light-soaking effects, while for NA device we noticed the rapid reduction of -13% during starting 400 seconds with the following saturation. Light-soaking (LS) effects are well known and widely studied especially for p-i-n PSCs[52]. The primary mechanism of the reduction of the output power under LS is related to the clustering of the ion defects (iodine[53], free A-site ions[54], interstitials[55] and vacancies[56]) at the interfaces with CTLs and trapping of the charge carriers. Considering the high concentration of the intrinsic point defects in IBS NiO, we relate the LS soaking effect of NA devices to the accelerated chemical reaction at NiO/perovskite interface through the bonding of non-coordinated metal cation sites of NiO[57] with A-site organic cation of the perovskite absorber as reported in Ref. [42].

To estimate the charge transport behavior, we measured the dark IV curves for the fabricated PSCs (**fig.5(d)**). The results showed the presence of strong hysteresis with shifts of minimum current from zero voltage. As reported in the literature for NiO-based devices[58], the dark hysteresis can be related to the electrochemical interaction at the HTL/perovskite[59] interface between ionic defects migration from the absorber (iodine vacancies, interstitials and antisites[60], [61]) and point defect in non-fully stoichiometric nickel oxide[62]. Additional information about the charge transport was obtained by analyzing the diode behavior of the fabricated PSCs with extraction of series resistance ($R_s$), shunt resistance ($R_{sh}$) and dark leakage current presented in the **tab.2**. The Rs values extracted from the dark IV curves at high voltage regions (1.0-1.2V) show the decrease from 5.036 Ohm for NA device to 0.839 Ohm for 300C PSC, which clearly highlight the improved contact properties after annealing post-treatment of HTL. This can be related to higher charge carrier mobility of 300C NiO (52.56 cm$^2$V$^{-1}$cm$^{-1}$) compared to NA; 150C, and improved energy level alignment, which also manifested in the higher values of FF measured for PSCs under light exposure. The calculated $R_{sh}$ values showed significant improvement with an increase from 10$^5$ Ohm for NA device to the order of 10$^8$ Ohm for 150C and 300C PSCs, respectively. The dark leakage current was reduced from ~10$^{-6}$ A/cm$^2$ for NA device to ~10$^{-7}$ A/cm$^2$. We assume that improvement of shunt properties originated from the morphology of the annealed NiO films with increased grain size of HTL and suppressed charge transport through highly conductive grain boundaries.

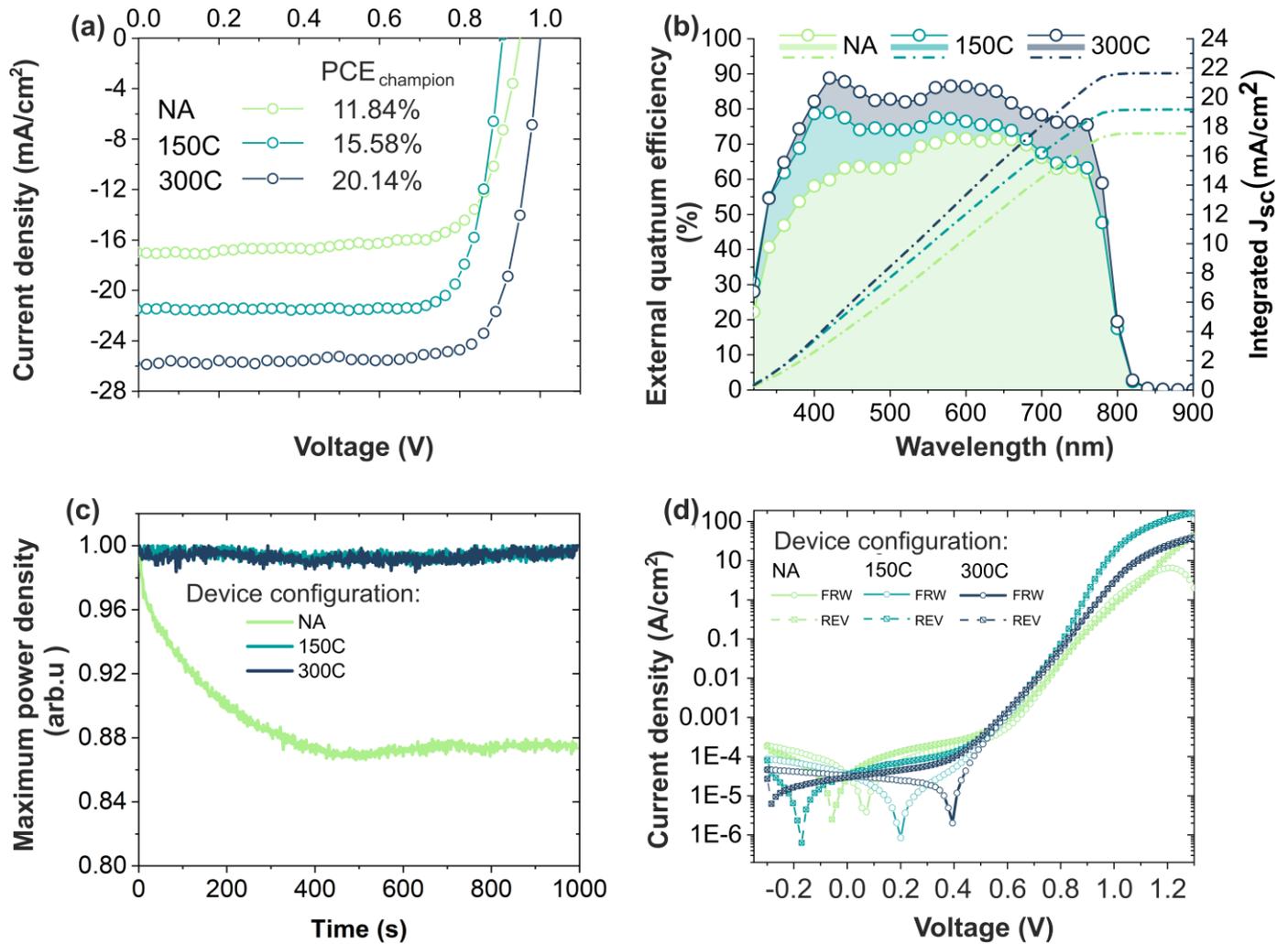

Figure 5 – The JV curves for the champion devices measured under standard conditions 1.5 AM G (a); the external quantum efficiency spectra for the investigated PSCs (b); the stabilization of the maximum power output under continuous light soaking for the NA, 150C and 300C devices (c); the dark JV curves for the investigated configurations of the PCSs

Table 2 – The extracted parameters after fitting of IV curves with 2-diode model and analysis of the dark JVs

| Device config. | Rs (Ohm) | Rsh (Ohm) | J$_{leakage}$ (mA/cm$^2$) |
|---|---|---|---|
| NA | 5.036 | 4.53E5 | 3E-3 |
| 150C | 1.602 | 1.58E8 | 9E-4 |
| 300C | 0.839 | 1.28E8 | 7E-4 |

To estimate the potential of IBS process for up-scaling and low-temperature fabrication, we manufactured perovskite solar modules on the substrates 100x100 mm$^2$ and flexible PSCs on PET (125 micron thick). The IV performance of the best-performing prototypes and the image of the devices are presented in the **fig.6**. The large area perovskite module with ion-beam sputtered NiO HTL consisted of 12 in-series connected sub-cells (4.58 cm$^2$ area for a single cell), which presents a reasonably high $V_{oc}$ of 11.35 V. The $I_{sc}$ of the module was 81.2 mA. Despite of small thickness of the NiO HTL (6.9 nm), the large area module showed the absence shunting or potential impact of micro pin-holes. The resulting PCE showed 8.91%, which was affected by high $R_s$. We fabricated a large area module using ultraviolet laser scribing procedure following *Castriotta et al.*[63], and assume that PCE of the modules fabricated with IBS NiO could be sufficiently improved with a proper optimization of P1-P3 multistep process. A flexible PSCs on PET substrates have been also fabricated with a temperature limit of 150 ˚C due to the thermal instability of such plastics[64]. The small area device on PET shows a reduced PCE compared to the PSCs manufactured on glass. The NA PSCs on PET has a PCE=9.92% (-16% relatively the device on glass). The 150C PSC on PET shows a drop of the performance down to 7.67% (-51% relatively the device on glass), mainly related to the bending of the substrate due to the high temperature of the NiO (see Ref.[65]) that also induce a deterioration of the interface quality between ITO, NiO and perovskite absorber. However, we found that the decrease of annealing temperature to 120 ˚C results in improvement of PCE up to 10.32%. The difference in the performance of flexible PSCs compared to the glass ones was mainly driven by changes of $J_{sc}$. The post-treatment annealing regime at 150ºC induces the warpage of the plastics substrate which affect to the interface quality in the thin-film device

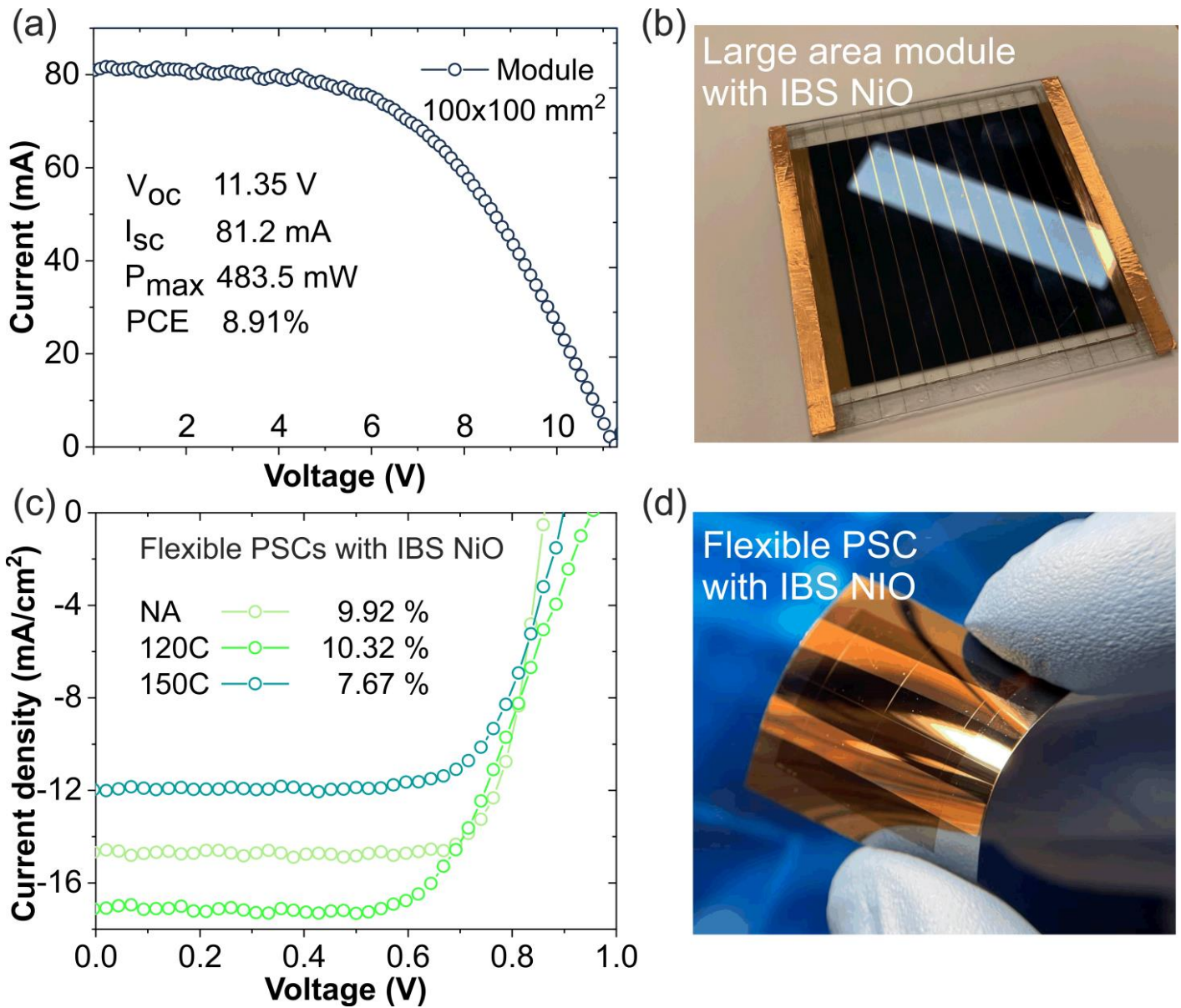

Figure 6 – The IV curve of the perovskite photomodule (active area 54.5 cm$^2$) with 300C NiO (a) the photo-image of the perovskite photomodule (b); the JV curve of the best performing flexible PSCs fabricated in the 125 micron-thick PET (c); the photoimage of the flexible PSC with IBS NiO HTL

**Conclusions:**

In conclusion, ion-beam sputtering has demonstrated considerable potential for the accurate fabrication of thin-film nickel oxide (NiO) as hole transport layers (HTLs) in perovskite solar cells (PSCs). Post-treatment air annealing at temperatures ranging from 150-300°C was determined to be crucial for enhancing the optical, surface, and electrical properties of the thin films. The bandgap ($E_g$) of the as-deposited NiO films was measured to be 3.28 eV, resulting in a relatively low transparency of 75-85% within the 350-450 nm wavelength region. Upon annealing at 300°C, the $E_g$ of NiO increased to 3.58 eV, leading to an improved transmittance of 88% in the short wavelength region and near-absence of absorption in the long wavelength region above 650 nm

The analysis of the structural and surface properties of NA NiO showed the thin-films have nanocrystalline morphology with $D_a$ of 9.8 nm and excess concentration of the oxygen (above stoichiometric), probably induced by presence of oxygen interstitials and Ni vacancies. The XRD analysis showed that NA samples are characterized by an increased unit cell volume 78.9425 Å$^3$ with respect to. The post-treatment air annealing at 300C provided an increase of the $D_a$ to 14.2 nm and significantly reduced the excess concentration of the oxygen in the composition. This led to a decrease in specific resistance from $10^{-4}$ to $10^{-5}$ Ohm*cm and an enhancement in Hall charge carrier mobility up to 52.56 cm$^2$V$^{-1}$s$^{-1}$. Utilizing a 7 nm ion-beam sputtered NiO as HTL for p-i-n PSCs exhibited promising output performance. Air annealing post-treatment yielded a champion power conversion efficiency (PCE) of 20.14% for 300°C devices, 15.58% for 150°C PSCs, and 11.84% for as-deposited samples. The PCE improvement mainly originated from enhanced short-circuit current density (Jsc) and open-circuit voltage (Voc). Dark current-voltage (IV) curve analysis revealed that 300°C NiO significantly reduced contact resistance and improved shunt properties, which could be attributed to a higher-quality NiO/perovskite interface with a reduced concentration of point defects.

The ion-beam sputtering technique demonstrated remarkable technological potential, enabling the fabrication of large-area modules with PCEs of 8.9% (54.5 cm$^2$ active area on a 10x10 cm$^2$ substrate) and flexible PSCs with a champion PCE of 10.32% when annealed at 120°C. This study highlights the extensive possibilities offered by ion-beam sputtering for the fabrication of charge transport layers in PSCs. The simplified fabrication of NiO HTLs using reactive sputtering of metal targets and air annealing exhibited competitive performance without the optimization typically required for oxide charge transport layers. Further enhancement of NiO properties for high-performance photovoltaics could be readily achieved using alloyed targets for doping (e.g., Ni-Cu or Ni-Co) or through multistep deposition with inorganic passivation interlayers such as MgO, NaCl, and others.

## ACKNOWLEDGEMENTS

Pavel Gostishchev, Lev Luchnikov and Danila Saranin gratefully acknowledge the financial support from the Ministry of Science and Higher Education of Russian Federation in terms of the Program Priority 2030 in NUST MISIS (grant №K2-2022-011).

## DATA AVAILABILITY

The data that supports the findings of this study are available in this article and its supplementary material.